\documentclass{PoS}

\usepackage{graphicx}
\usepackage{subcaption}

\newcommand{\xm}{\relax\ifmmode X_{\mathrm{max}} \else
  $X_{\mathrm{max}}$\fi}
\newcommand{\mxm}{\relax\ifmmode \left<X_{\mathrm{max}}\right> \else
  $\left<X_{\mathrm{max}}\right>$\fi}
\newcommand{\sxm}{\relax\ifmmode \sigma(X_{\mathrm{max}}) \else
  $\sigma(X_{\mathrm{max}})$\fi}
\newcommand{\nm}{\relax\ifmmode N_{\mathrm{max}} \else
  $N_{\mathrm{max}}$\fi}

\title{Telescope Array 10 Year Composition}

\ShortTitle{Telescope Array 10 Year Composition}

\author{\speaker{William Hanlon} for the Telescope Array
  Collaboration\footnote{For collaboration list see PoS(ICRC2019)1177}\\
        University of Utah, Department of Physics and Astronomy \& High
        Energy Astrophysics Institute, 201 James Fletcher Bldg., 115 S
        1400 E, Salt Lake City, UT 84112, USA\\
        E-mail: \email{whanlon@cosmic.utah.edu}}

\abstract{Estimates of the composition of ultra high energy cosmic
  rays (UHECRs) can be inferred by recording the depth of air shower
  maximum, \xm, for many showers and comparing the distributions to
  those predicted by Monte Carlo simulations. Traditionally, UHECR
  composition has relied upon comparison of the first and second
  moments of the \xm{} distributions to estimate the compatibility
  between data and simulations, but with the large UHECR datasets
  being built the current generation experiments better tests which
  compare full distributions can be employed. Such tests can be used
  to understand the accuracy with which UHECR composition can actually
  be understood at the current level of statistics and quantitatively
  measure the significance of agreement or disagreement with models in
  order to reject them. In this paper we present the most recent
  results of 10 years of Telescope Array hybrid \xm{}
  measurements which is found to agree with a predominantly light
  composition. In previously published results we have demonstrated
  the agreement of Telescope Array hybrid \xm{} data with single
  element models using systematic shifting of the data in order to
  ensure the shapes of the distributions are being compared. Here we
  present multi-component source models fit to hybrid \xm{} data and
  report on the relative fractions of those sources that best fit the
  data. Below $10^{19.1}$~eV TA hybrid data is found to be compatible
  with mixtures composed of predominantly light elements such as
  protons and helium.}

\FullConference{36th International Cosmic Ray Conference -ICRC2019-\\
		July 24th - August 1st, 2019\\
		Madison, WI, U.S.A.}

\begin{document}

\section{Introduction}\label{sec:intro}
Telescope Array (TA) is a large hybrid cosmic ray observatory located
in Millard County, Utah ($39.3^\circ$~N, $112.9^\circ$~W, 1400~m asl)
designed to observe ultra high energy cosmic rays with energies in
excess of $10^{18}$~eV. TA utilizes 507 plastic scintillation
counters, also referred to as surface detectors (SDs), placed over
$~700$~km$^2$ and 36 fluorescence detector (FD) telescopes,
distributed among three FD stations, to measure the energy, depth of
air shower maximum (\xm), and arrival direction of UHECRs. Refer to
\cite{AbuZayyad:2012kk,AbuZayyad:2000uu,Tameda:2009zza,Tokuno:2012mi}
for further details of TA's SD and FD operations.

\section{TA Ten Year \xm{} data}\label{sec:ta_xmax_data}
Hybrid reconstruction is done by searching for coincident events in
the SD and FD data streams that occur within 500$\mu$s. The
timing and geometry of the event from the SD event data is used to
constrain the location of the shower core on the ground, which greatly
improves the determination of the shower track in the FD
shower-detector plane. Using the improved geometry fit, the light
profile of the shower is fit using the FD information, providing
accurate measurements of energy and \xm{} that are better than
monocular FD reconstruction alone. Uncertainties in angular
quantities important to reconstruction of the shower track improve to
less than a degree, and relative uncertainties in distances improve to
less than 1\% when performing hybrid reconstruction.

\begin{figure}
  \centering
  \begin{subfigure}{0.45\linewidth}
  \includegraphics[clip,width=\textwidth]{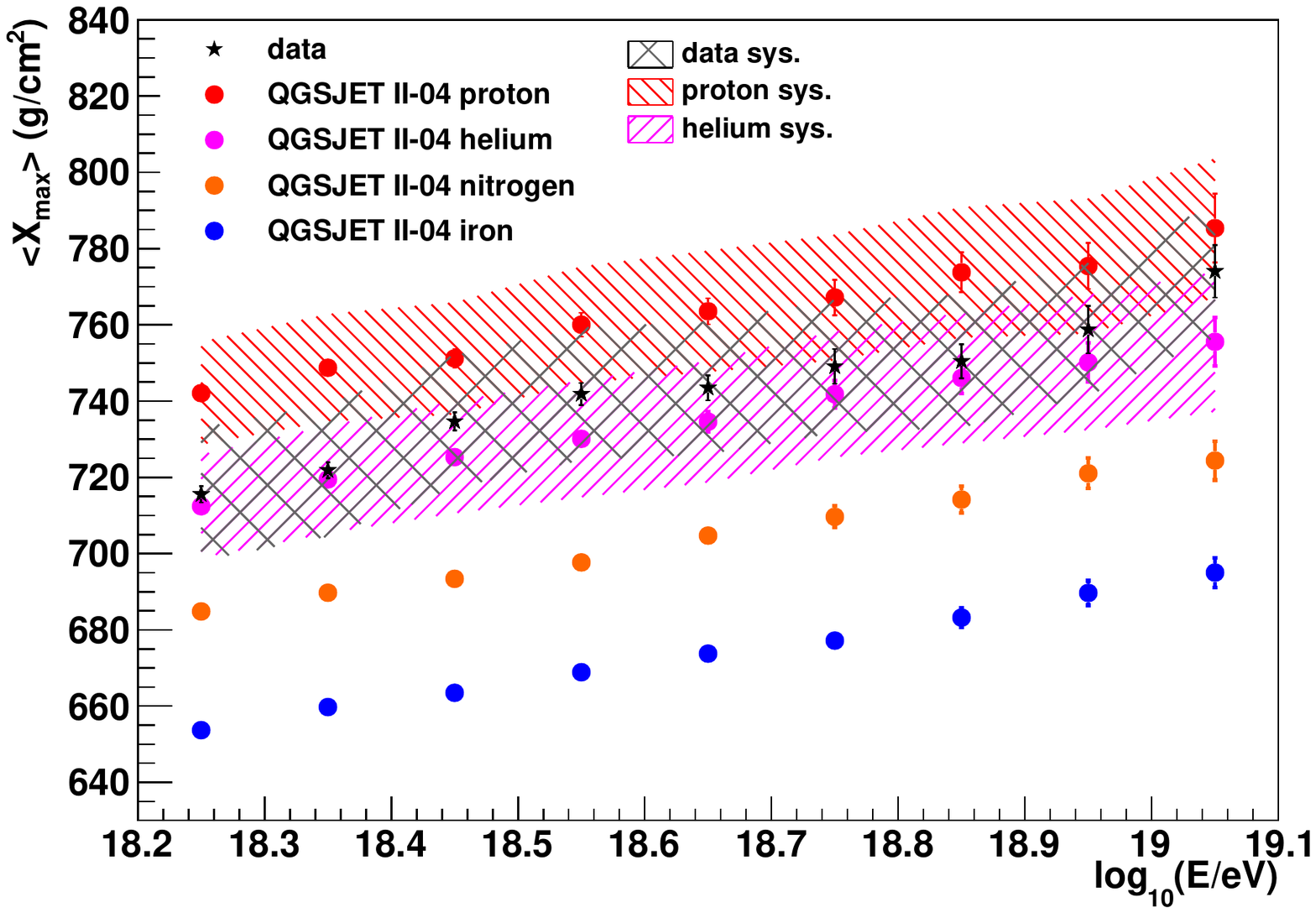}
  \caption{TA 10 year hybrid \mxm{}}
  \label{fig:brlr_mxm}
  \end{subfigure}%
  \qquad%
  \begin{subfigure}{0.45\linewidth}
      \includegraphics[clip,width=\textwidth]{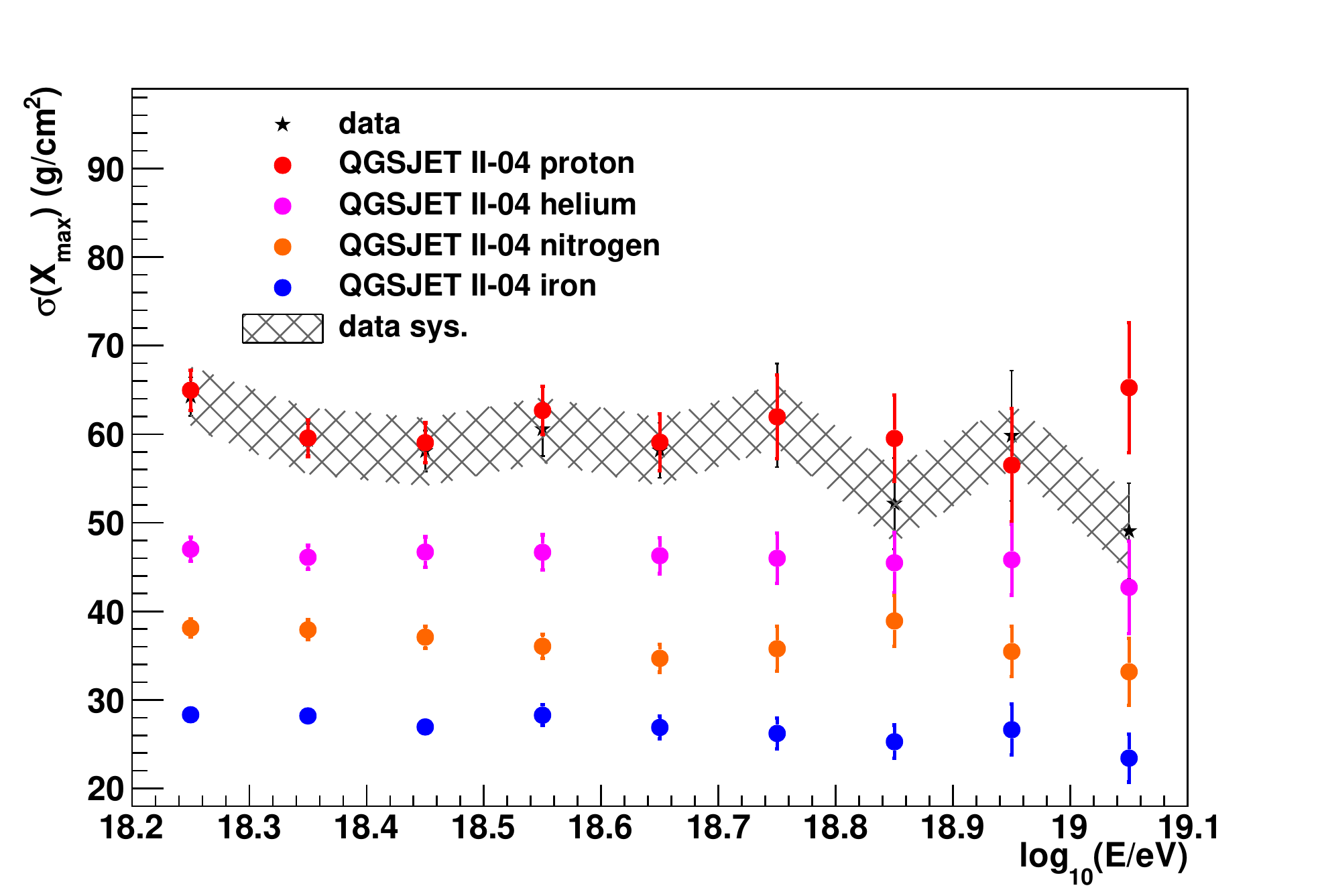}
      \caption{TA 10 year hybrid \sxm{}}
      \label{fig:brlr_sxm}
  \end{subfigure}
  \caption{Ten year TA hybrid \mxm{} and \sxm{} measurements using
    Black Rock Mesa and Long Ridge fluorescence detectors and the
    surface detector array.}
  \label{fig:brlr_mxm_and_sxm}
\end{figure}

TA's highest statistics measure of composition is done using Black
Rock Mesa and Long Ridge (BR/LR) hybrid. Events that trigger the BR or
LR FD stations are time matched to events that also trigger the SD
array. If an event is observed by both FD stations, the shower
parameters from the site with the better hybrid shower profile
$\chi^{2}$ is chosen. Data and Monte Carlo are processed via the same
analysis software and the same quality cuts are applied: the event
core must be well within the SD array more than 100~m from the
boundary, FD track length $10^\circ$ or greater, 11 or more good tubes
recorded by FDs, shower-detector plane angle ($\psi$) less than
$130^\circ$, time extent of the FD track greater than 7~$\mu$s, zenith
angle less than $55^\circ$, \xm{} must be observed within the field of
view of FDs, and weather cuts to ensure atmospheric quality is
good. 3560 events, collected over the period 27 May 2008 to 28
November 2017, were selected after application of these cuts. The
systematic uncertainty on the \mxm{} data is 17~g/cm$^{2}$ (black band
in the figure). \xm{} bias and resolution are < 1 and 17.2~g/cm$^2$
respectively, and energy resolution is
5.7\%~\cite{Abbasi:2018nun}. The absolute FD energy scale energy
uncertainty is
21\%~\cite{Abu-Zayyad:2013jra,Abu-Zayyad:2013qwa}. Figure~\ref{fig:brlr_mxm_and_sxm}
shows the observed \mxm{} and \sxm{} for 10 years of data between
$10^{18.2}$ and $10^{19.1}$~eV along with predictions of QGSJET~II-04
proton, helium, nitrogen, and iron .

\section{Single Element UHECR Composition}\label{sec:single_element}
A more rigorous test of the compatibility of TA data with models is to
test the entire distributions instead of \mxm{} and \sxm{}. We test
the agreement of TA data and single element models in this case by
computing the maximum likelihood of the data and Monte Carlo
distributions and including a systematic shift of the data to account
of systematic uncertainty in \xm{} either in our analysis, the models
being tested, or both. This tests the shapes of the distributions
which contain a good deal of information because of the exponential
tail of the light components. For each component and in each energy
bin, the data is shifted and the shift with the largest maximum
likelihood is recorded. Then the probability ($p$-value) of observing
a likelihood at least as extreme as found between the data and the
model is calculated~\cite{Abbasi:2018nun}.

\begin{figure}
    \centering
    \begin{subfigure}{0.45\linewidth}
      \includegraphics[clip,width=\textwidth]{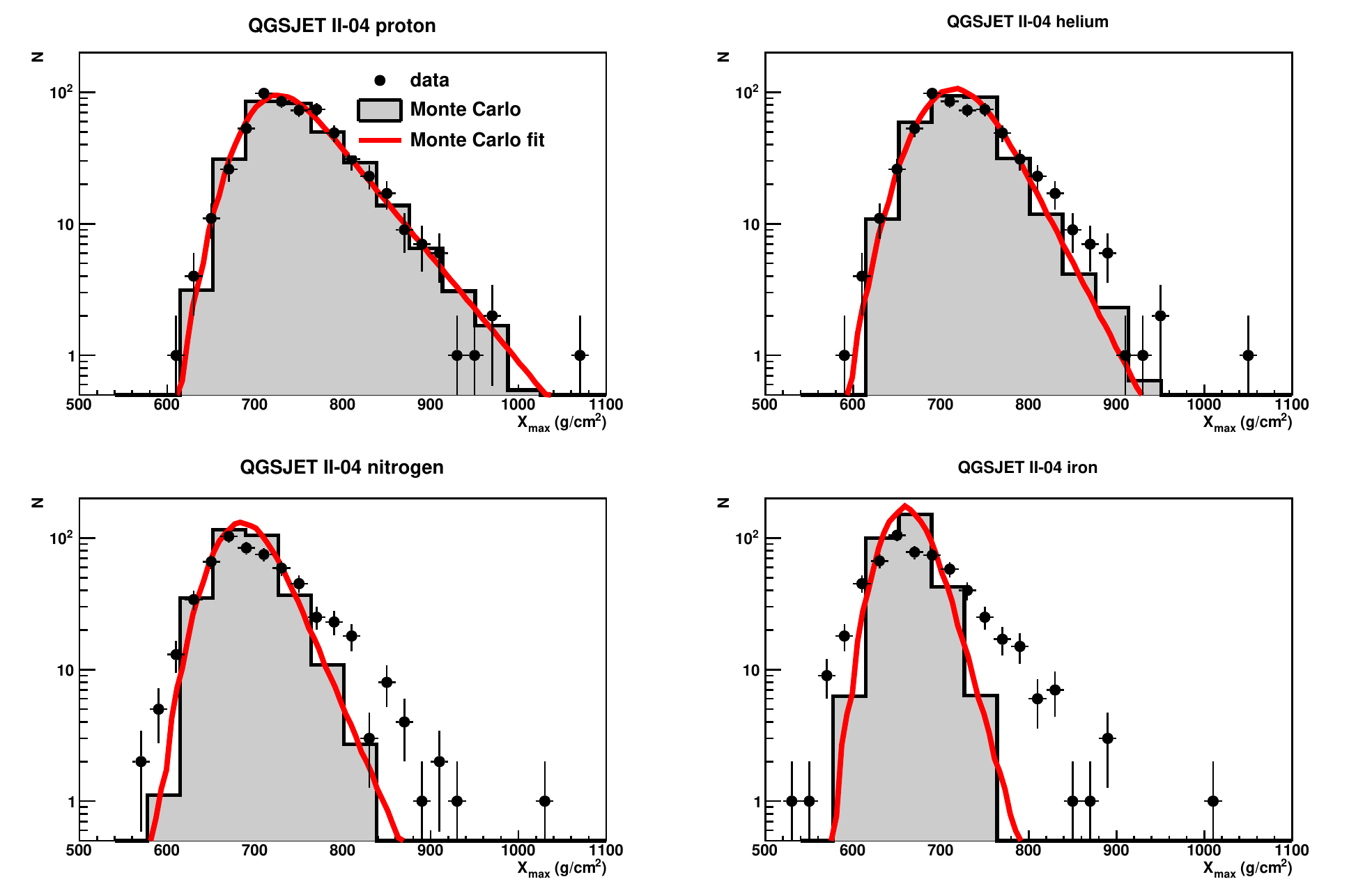}
      \caption{Data/MC comparison of four elements in one energy bin.}
      \label{fig:single_element_comparisons}
    \end{subfigure}%
    \qquad%
    \begin{subfigure}{0.45\linewidth}
      \includegraphics[clip,width=\textwidth]{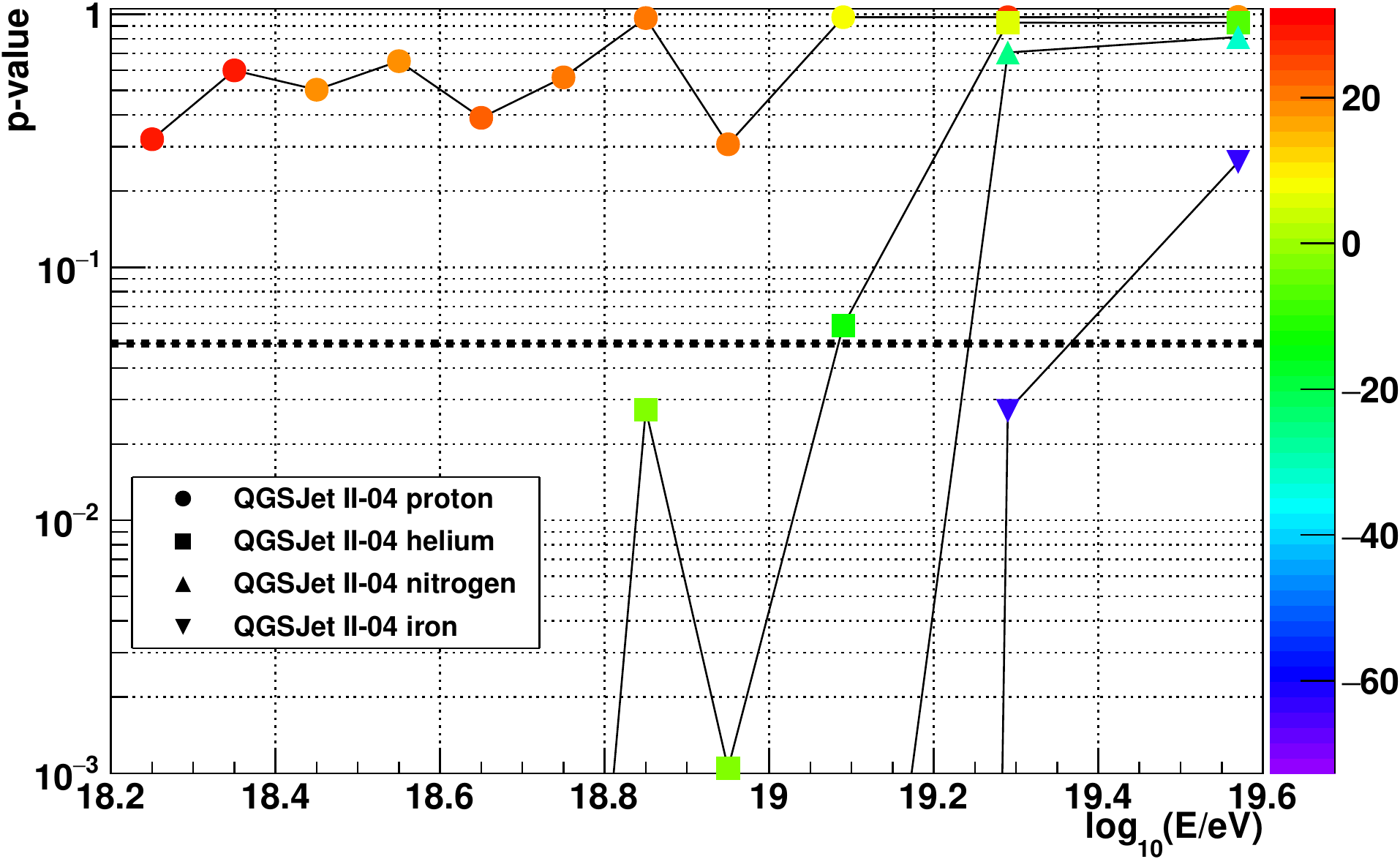}
      \caption{Data/MC single element compatibility}
      \label{fig:single_element_compatibility}
    \end{subfigure}
    \caption{Comparison of TA hybrid \xm{} distributions to single
        element predictions of QGSJET II-04 proton, helium, nitrogen,
        and iron after systematic shifting  and compatibility
        with different simulated elements.}
    \label{fig:single_element_test}
\end{figure}

Figure~\ref{fig:single_element_comparisons} shows the data and Monte
Carlo predictions of QGSJET~II-04 proton, helium, nitrogen, and iron
for one energy bin under this test. Notice that proton and helium appear to
agree with the data well, especially in the tails of the
distributions, whereas nitrogen and iron do not resemble the data. The
same data is shown in each panel, but is systematically shifted as
described above. Figure~\ref{fig:single_element_compatibility}
summarizes these tests showing the $p$-values obtained and the shifts
required to maximize the likelihood. If the $p$-value is less than
0.05, we say that the data is not compatible with the element in
question for that energy bin. As the figure shows, TA hybrid \xm{}
data is compatible with QGSJET II-04 proton from $10^{18.2}$ to
$10^{19.9}$~eV with systematic shifting of about 20~g/cm$^2$
needed. Other elements are not compatible with the data until
$10^{19}$~eV. In the last energy bin, all four single element tests
indicate compatibility with the data. This is due to low statistics in
that bin, where only 19 events are recorded. Notice also that
iron requires a shift of about 50~g/cm$^2$, which is much larger than
our systematic uncertainty.

\section{Multiple Element UHECR
  Composition}\label{sec:multiple_element} 
\begin{figure}
  \centering
  \includegraphics[clip,width=0.85\textwidth]{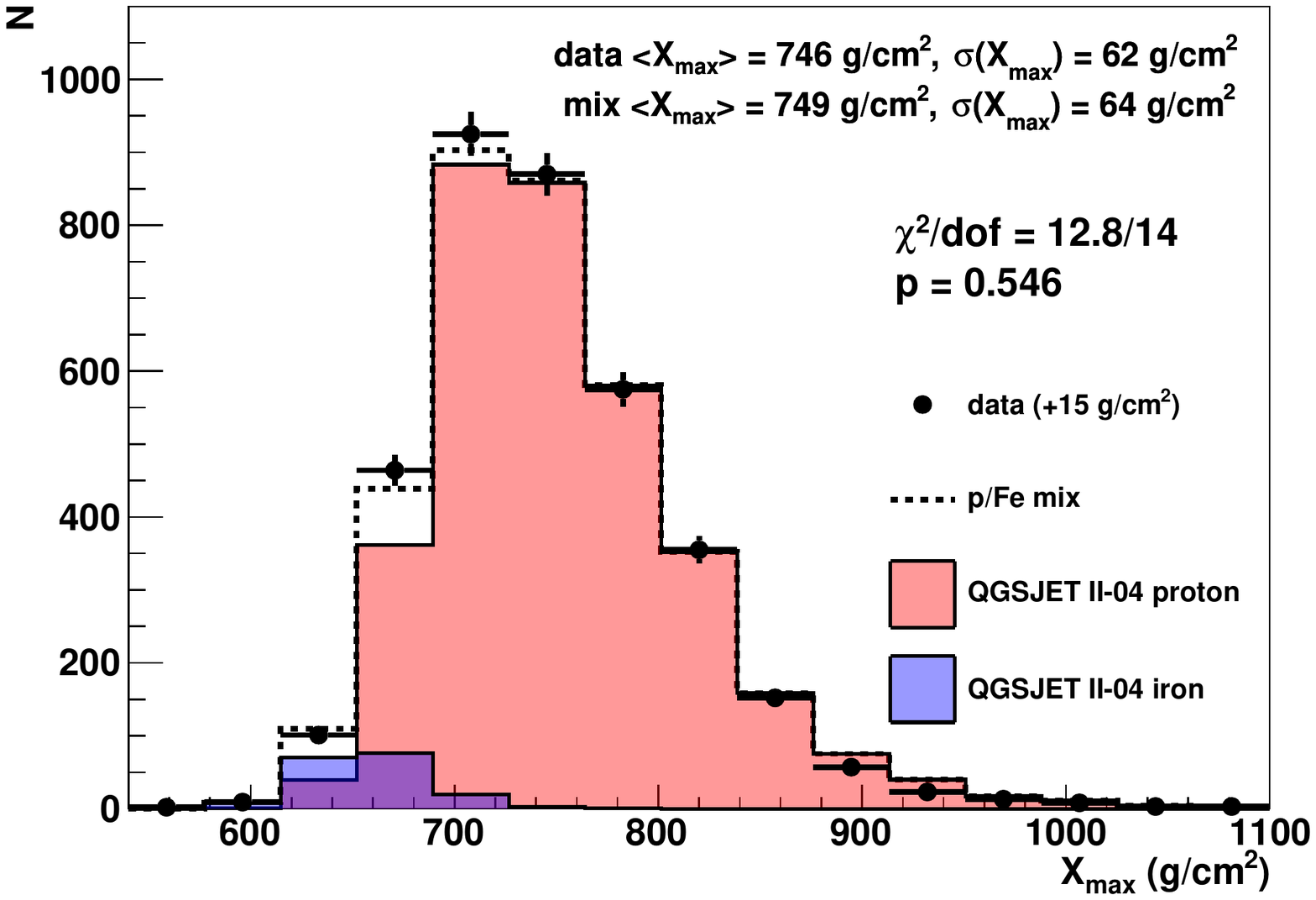}
  \caption{TA hybrid \xm{} compared to QGSJET~II-04 proton and iron
    model}
  \label{fig:p_fe_model_dist}
\end{figure}

\begin{figure}
    \centering
    \begin{subfigure}{0.45\linewidth}
      \includegraphics[clip,width=\textwidth]{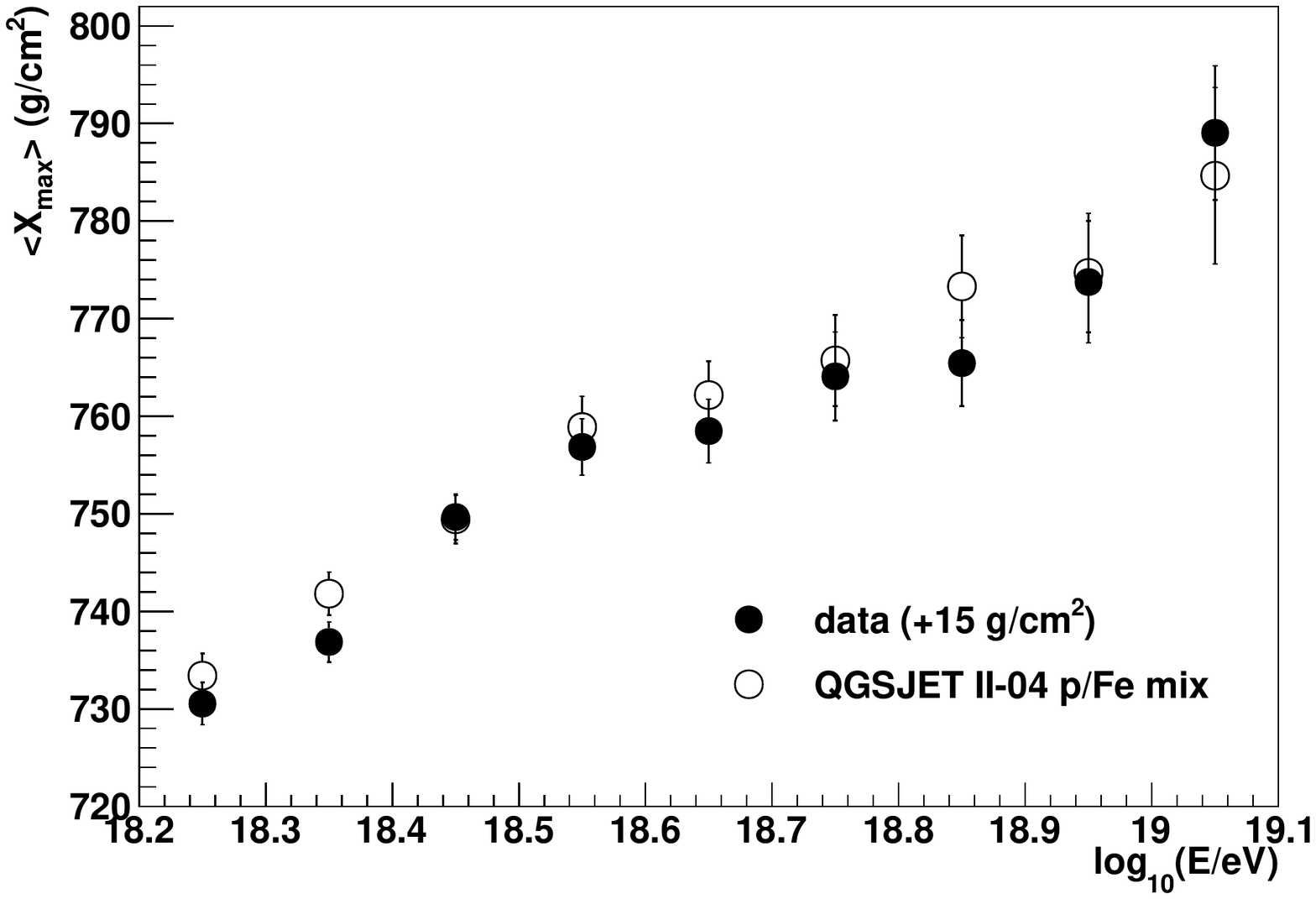}
      \caption{\mxm{} of TA hybrid data and QGSJET~II-04 proton-iron mix}
      \label{fig:2comp_pi_mxm}
    \end{subfigure}%
    \qquad%
    \begin{subfigure}{0.45\linewidth}
      \includegraphics[clip,width=\textwidth]{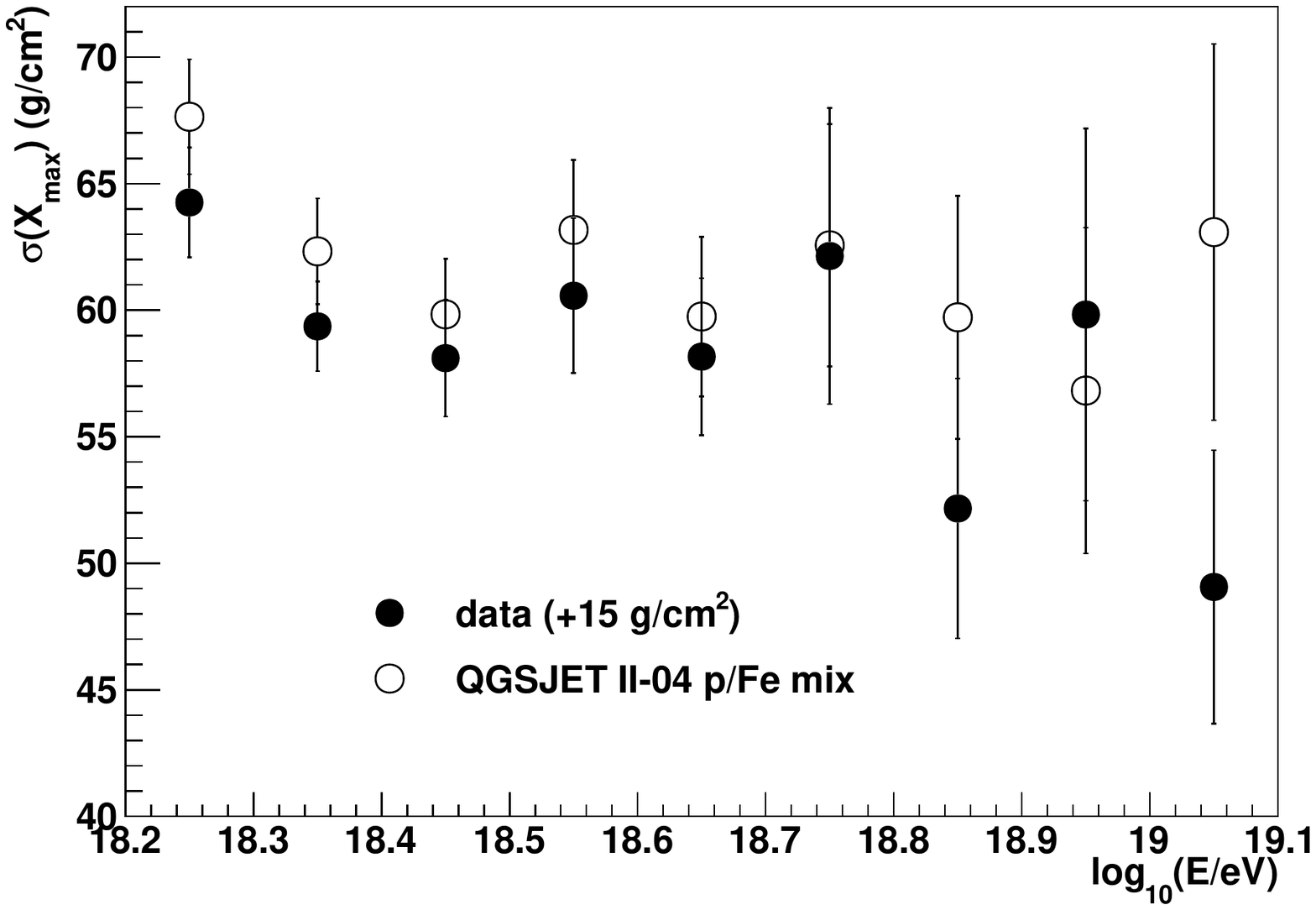}
      \caption{\sxm{} of TA hybrid data and QGSJET~II-04 proton-iron mix}
      \label{fig:2comp_pi_sxm}
    \end{subfigure}
    \caption{Comparison of \mxm{} and \sxm{} of TA hybrid data and a
      QGSJET~II-04 proton and iron mixture.}
    \label{fig:2comp_pi_comparisons}
\end{figure}

We can extend this type of
test to mixtures of elements. The simplest assumption is that UHECRs
are comprised of a light and heavy component, e.g., proton and
iron. Using the method outlined by Barlow \&
Beeston~\cite{Barlow:1993dm} and implemented in
ROOT~\cite{Brun:1997pa} as the TFractionFitter class, we find the
source weights (fractions) of reconstructed QGSJET~II-04 proton and
iron which best fit the data in energy bins between $10^{18.2}$ -
$10^{19.1}$~eV. We can also scan for the minimum $\chi^2$ of these
fits as we systematically shift the data within our quoted \xm{}
systematics. Figure~\ref{fig:p_fe_model_dist} shows the \xm{}
distribution of the data, mix, and contributions of proton and
iron. The minimum $\chi^2$ for this model is found by shifting up
by $+15$~g/cm$^2$. The fractions of proton and iron found to best fit
the data is 95\% proton and 5\% iron, with good agreement in the means
and widths of the data and mix distributions. The \mxm{} of the data
and proton-iron mix are 746~g/cm$^2$ and 749~g/cm$^2$ respectively and
the \sxm are 62~g/cm$^2$ and 64~g/cm$^2$. The \mxm{} and \sxm{} of the
data and the proton-iron mix as a function of energy are shown in
figure~\ref{fig:2comp_pi_comparisons}.

\begin{figure}
  \centering
  \includegraphics[clip,width=0.85\textwidth]{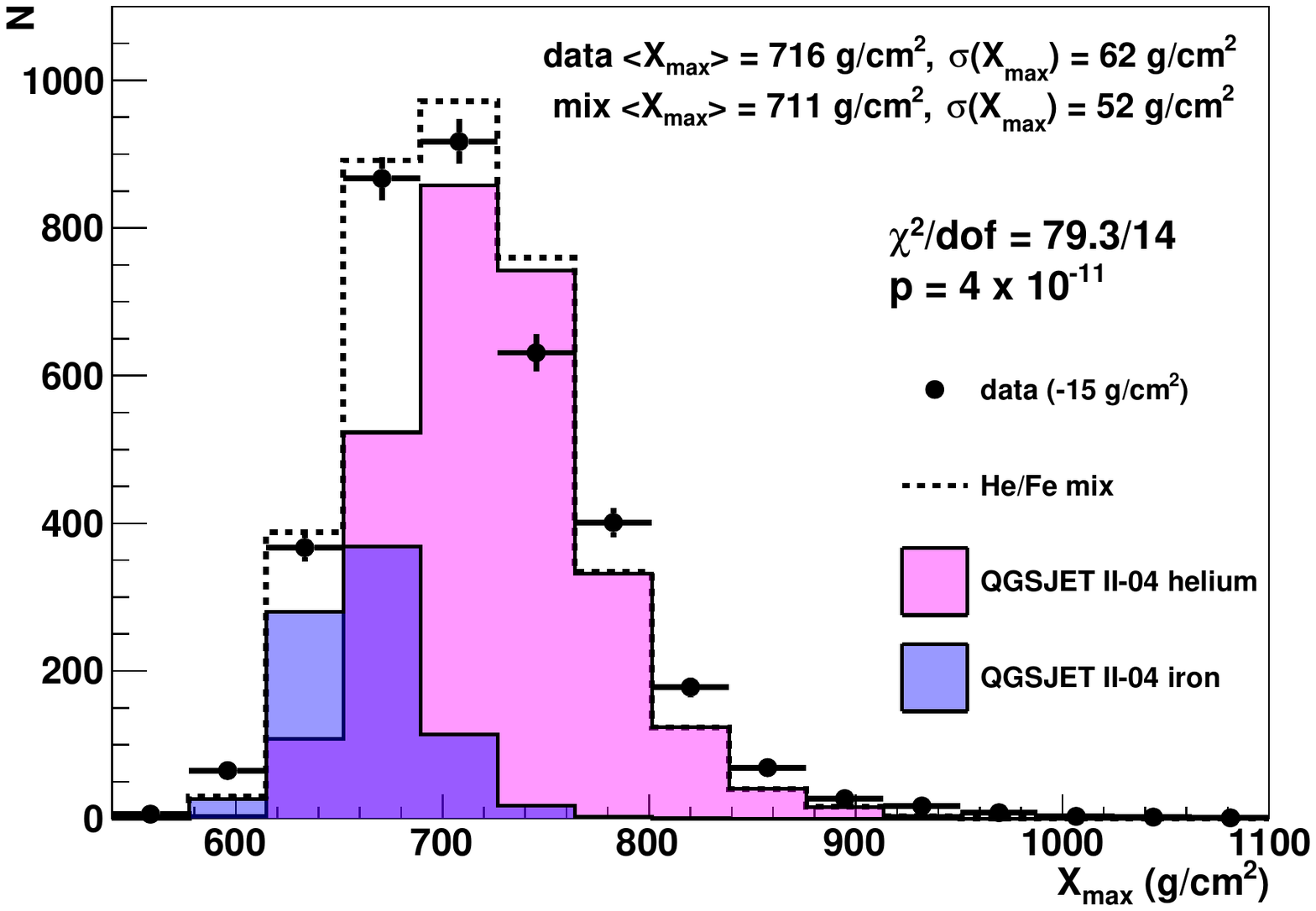}
  \caption{TA hybrid \xm{} compared to QGSJET~II-04 helium and iron
    model}
  \label{fig:he_fe_model_dist}
\end{figure}

\begin{figure}
    \centering
    \begin{subfigure}{0.45\linewidth}
      \includegraphics[clip,width=\textwidth]{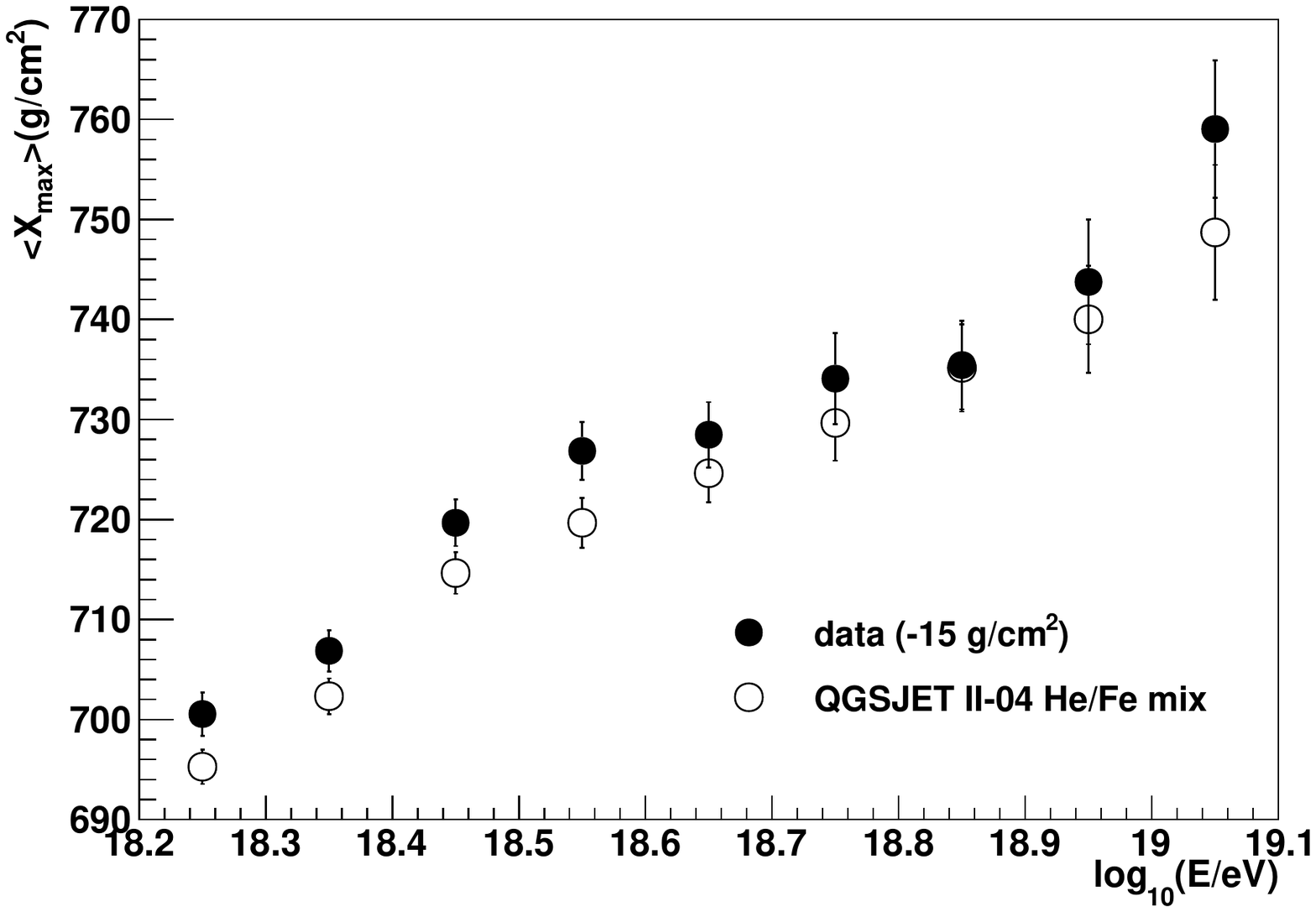}
      \caption{\mxm{} of TA hybrid data and QGSJET~II-04 helium-iron mix}
      \label{fig:2comp_hi_mxm}
    \end{subfigure}%
    \qquad%
    \begin{subfigure}{0.45\linewidth}
      \includegraphics[clip,width=\textwidth]{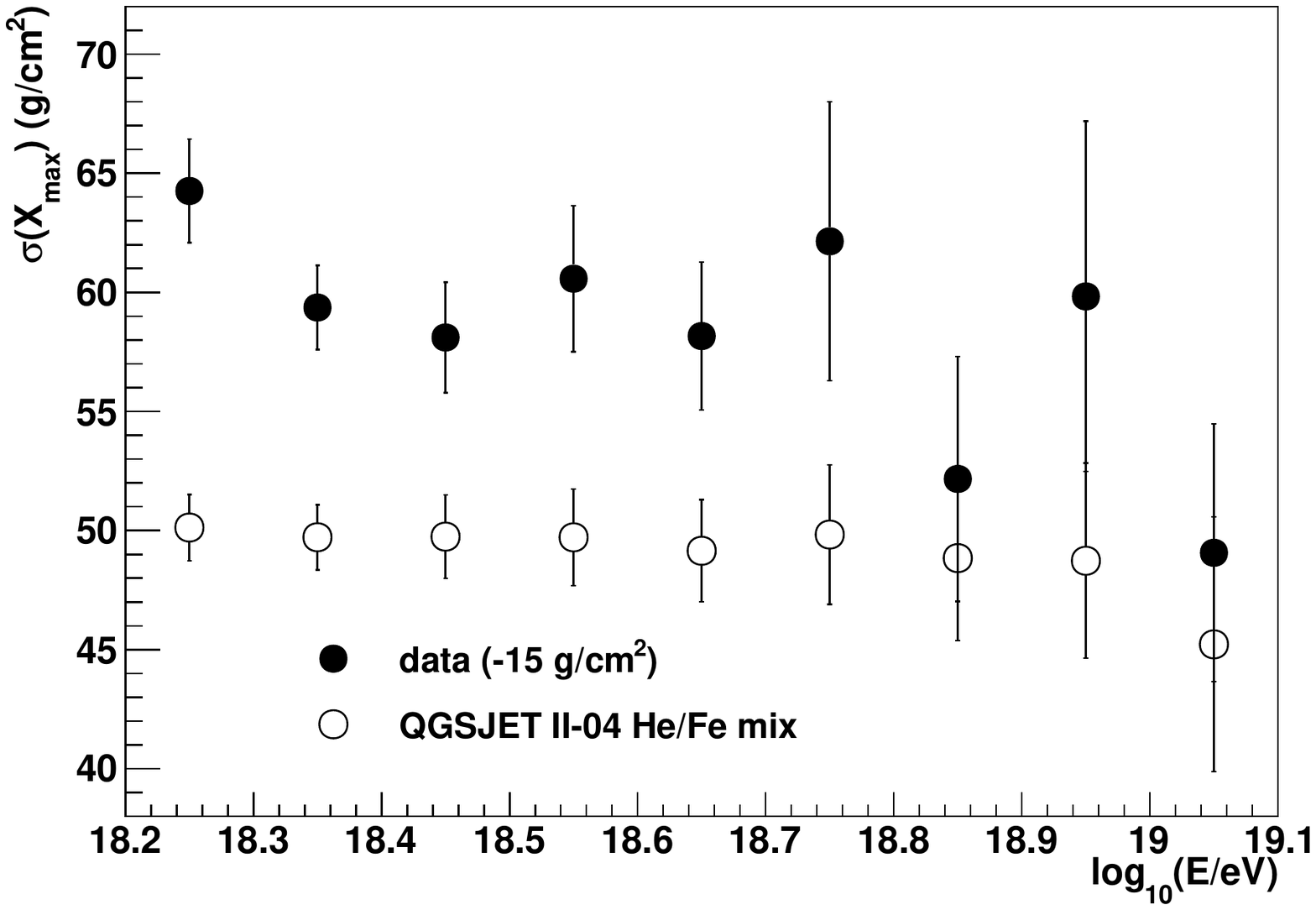}
      \caption{\sxm{} of TA hybrid data and QGSJET~II-04 helium-iron mix}
      \label{fig:2comp_hi_sxm}
    \end{subfigure}
    \caption{Comparison of \mxm{} and \sxm{} of TA hybrid data and a
      QGSJET~II-04 helium and iron mixture.}
    \label{fig:2comp_hi_comparisons}
\end{figure}

It is instructive to look at the next simplest light/heavy model:
helium and iron. Figure~\ref{fig:he_fe_model_dist} shows the \xm{}
distributions of this two component model and the data and
figure~\ref{fig:2comp_hi_comparisons} shows the \mxm{} and \sxm{} of
such a mixture. This composition mix required a -15~g/cm$^2$ shift in
the data to find the minimum $\chi^2$ and results in a mixture of 77\%
helium and 23\% iron. As figure~\ref{fig:2comp_hi_comparisons} shows,
while the \mxm{} of the mix is within statistical uncertainties of the
shifted data, \sxm{} does not resemble the data at all. It's important
to recall that uniform systematic shifting of a distribution does not
change \sxm{}. Because protons contribute prominently to the tail of
\xm{} distributions, this result indicates that some protons must be
present in any mixture we create. The difference in the tails of the
data and the helium-iron mixture can be clearly seen in the right part
of the distributions shown in figure~\ref{fig:he_fe_model_dist}. The
simple helium-iron can be used to measure a lower bound on the proton
content of \xm{} distributions.

\begin{figure}
  \centering
  \includegraphics[clip,width=0.85\textwidth]{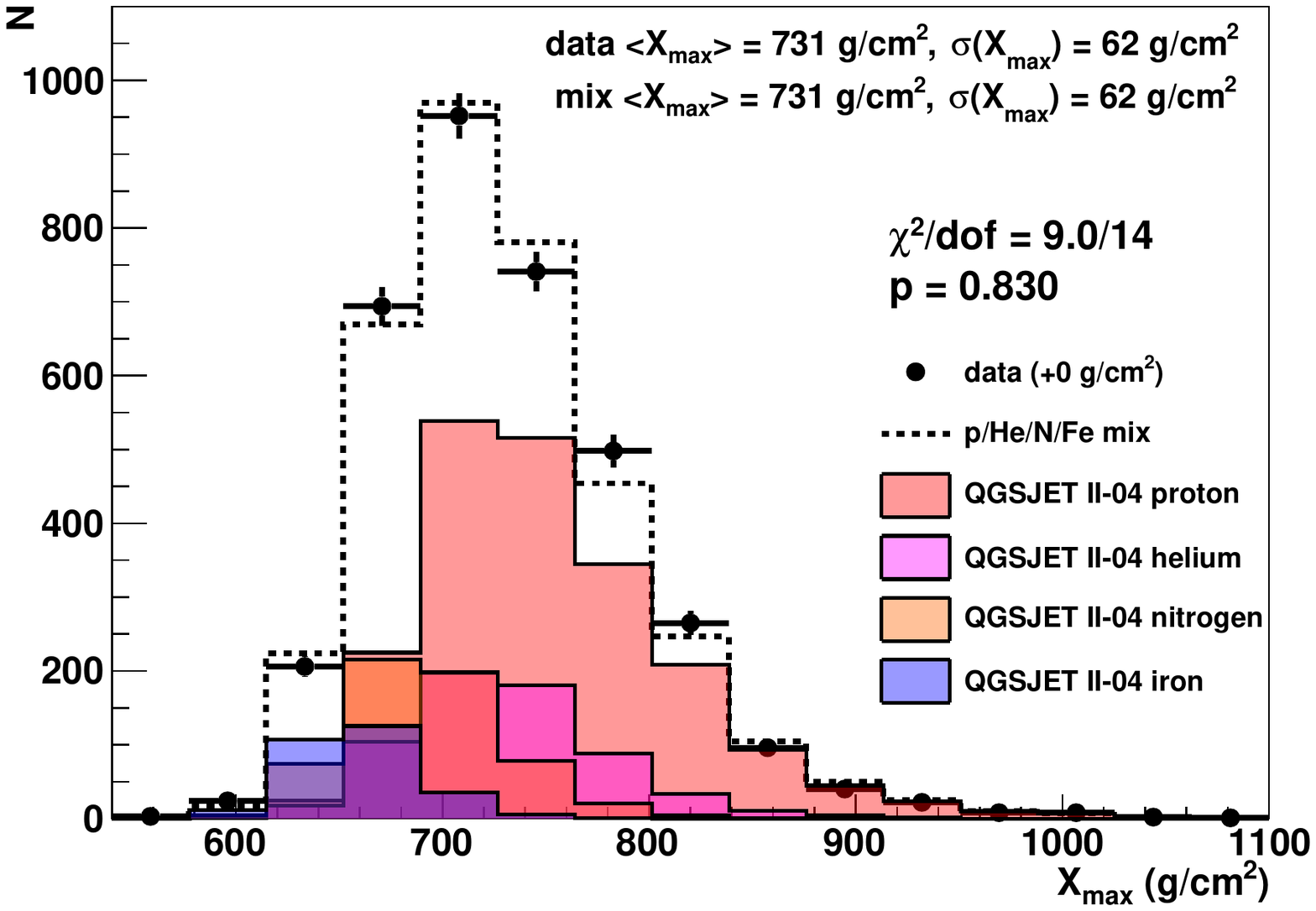}
  \caption{TA hybrid \xm{} compared to QGSJET~II-04 four component model.}
  \label{fig:4comp_model_dist}
\end{figure}

\begin{figure}
    \centering
    \begin{subfigure}{0.45\linewidth}
      \includegraphics[clip,width=\textwidth]{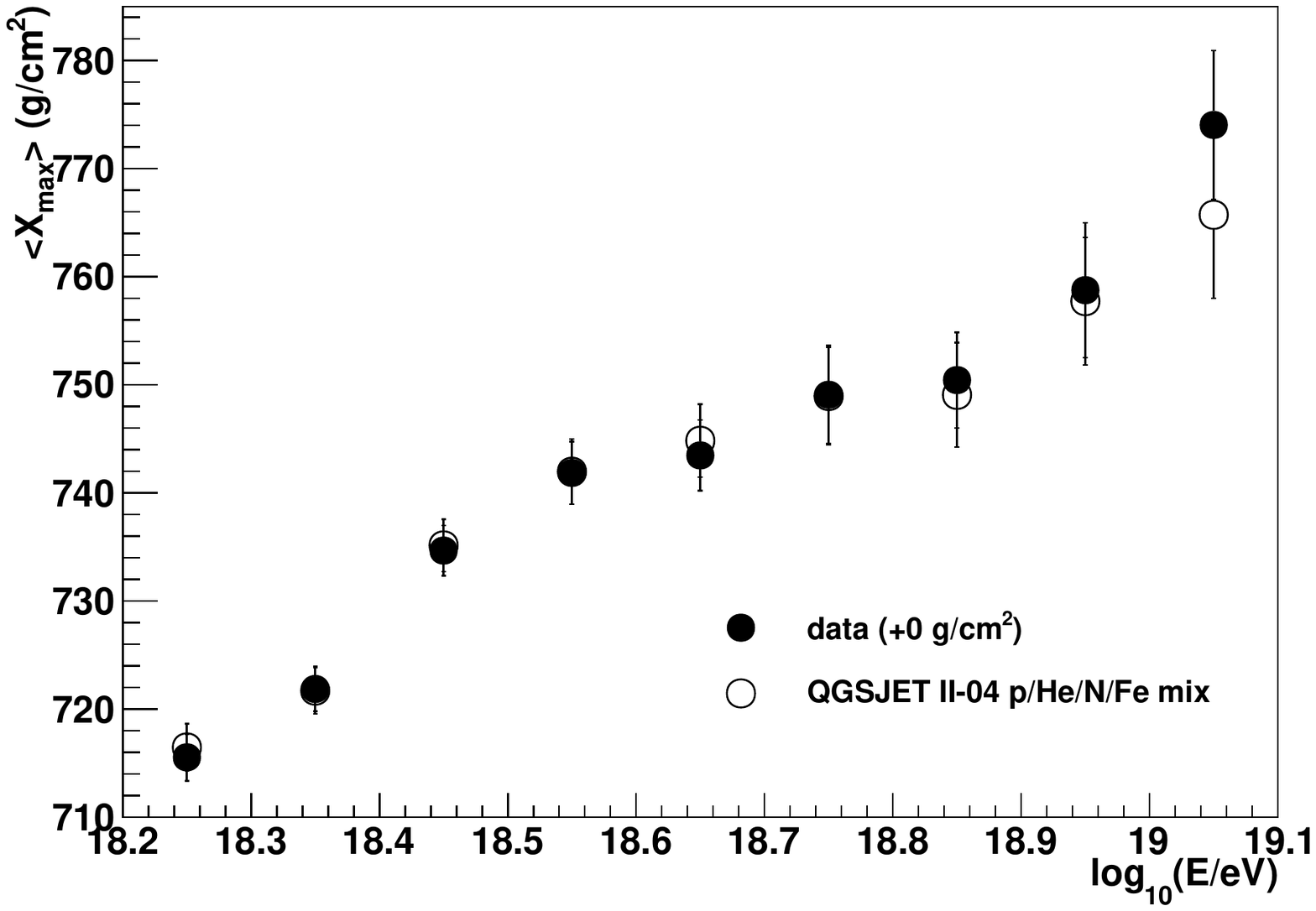}
      \caption{\mxm{} of TA hybrid data and QGSJET~II-04 four
        component mix.}
      \label{fig:4comp_mxm}
    \end{subfigure}%
    \qquad%
    \begin{subfigure}{0.45\linewidth}
      \includegraphics[clip,width=\textwidth]{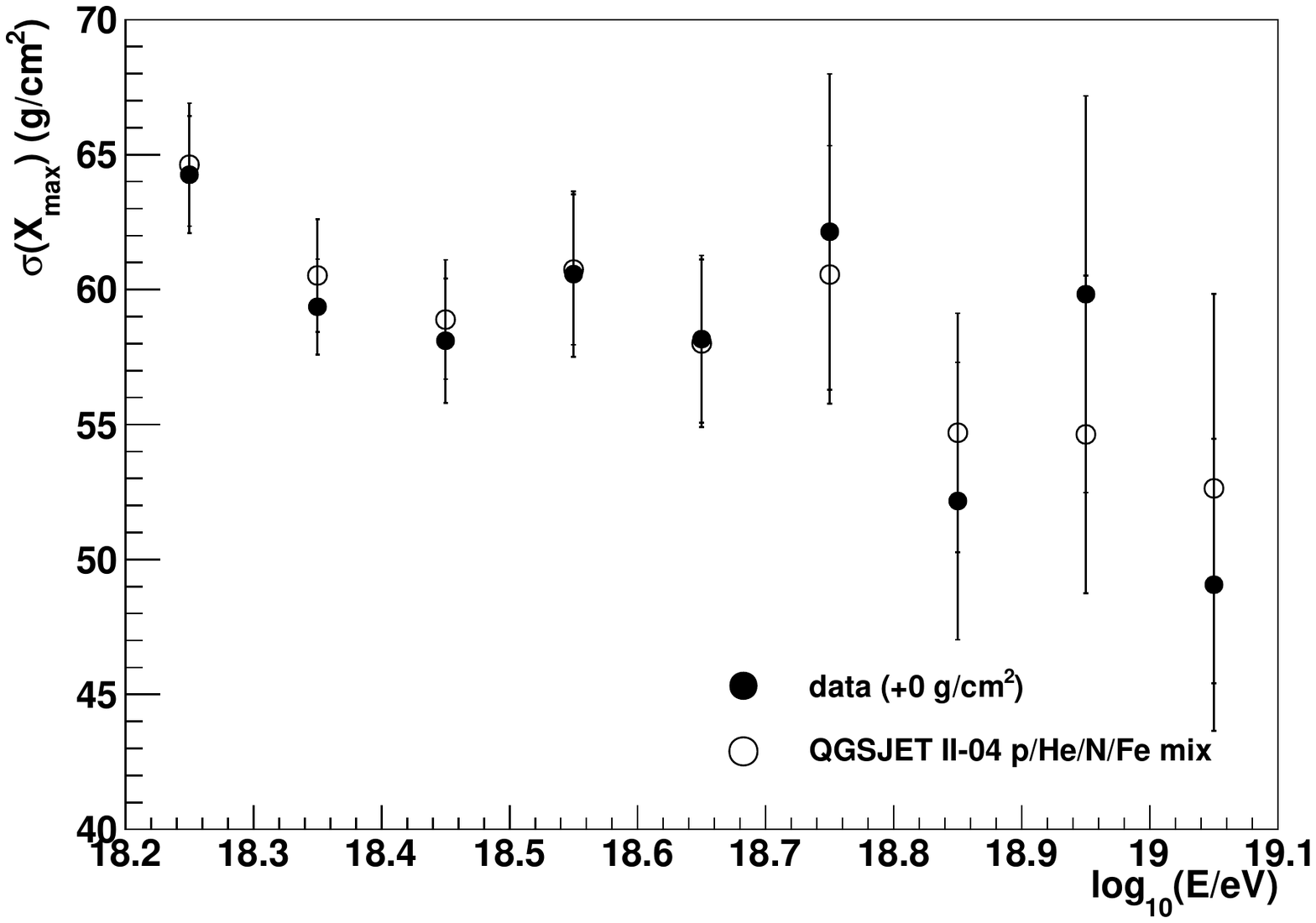}
      \caption{\sxm{} of TA hybrid data and QGSJET~II-04 four
        component mix.}
      \label{fig:4comp_sxm}
    \end{subfigure}
    \caption{Comparison of \mxm{} and \sxm{} of TA hybrid data and a
      QGSJET~II-04 four component mixture.}
    \label{fig:4comp_comparisons}
\end{figure}

The Auger collaboration has published a similar analysis in which they
fit their data to a four component model~\cite{Aab:2014aea}. We now
perform the same analysis using the QGSJET~II-04 hadronic
model. Figure~\ref{fig:4comp_model_dist} shows the \xm{} distributions
of the data and model for the energy range $10^{18.2}$ -
$10^{19.1}$~eV under the assumption of a four component mixture. Note
that no systematic shift is applied to the data. \mxm{} and \sxm{} of
the data and mix, shown in energy bins in figures~\ref{fig:4comp_mxm} and
\ref{fig:4comp_sxm}, agree well and the $\chi^2$/dof of the
distributions is 9.0/14. The mix consists of 57\%, 18\%, 17\%, 8\%
proton, helium, nitrogen, and iron respectively. 75\% of this
mixture are the light elements proton and helium.

\begin{figure}
  \centering
  \includegraphics[clip,width=0.85\textwidth]{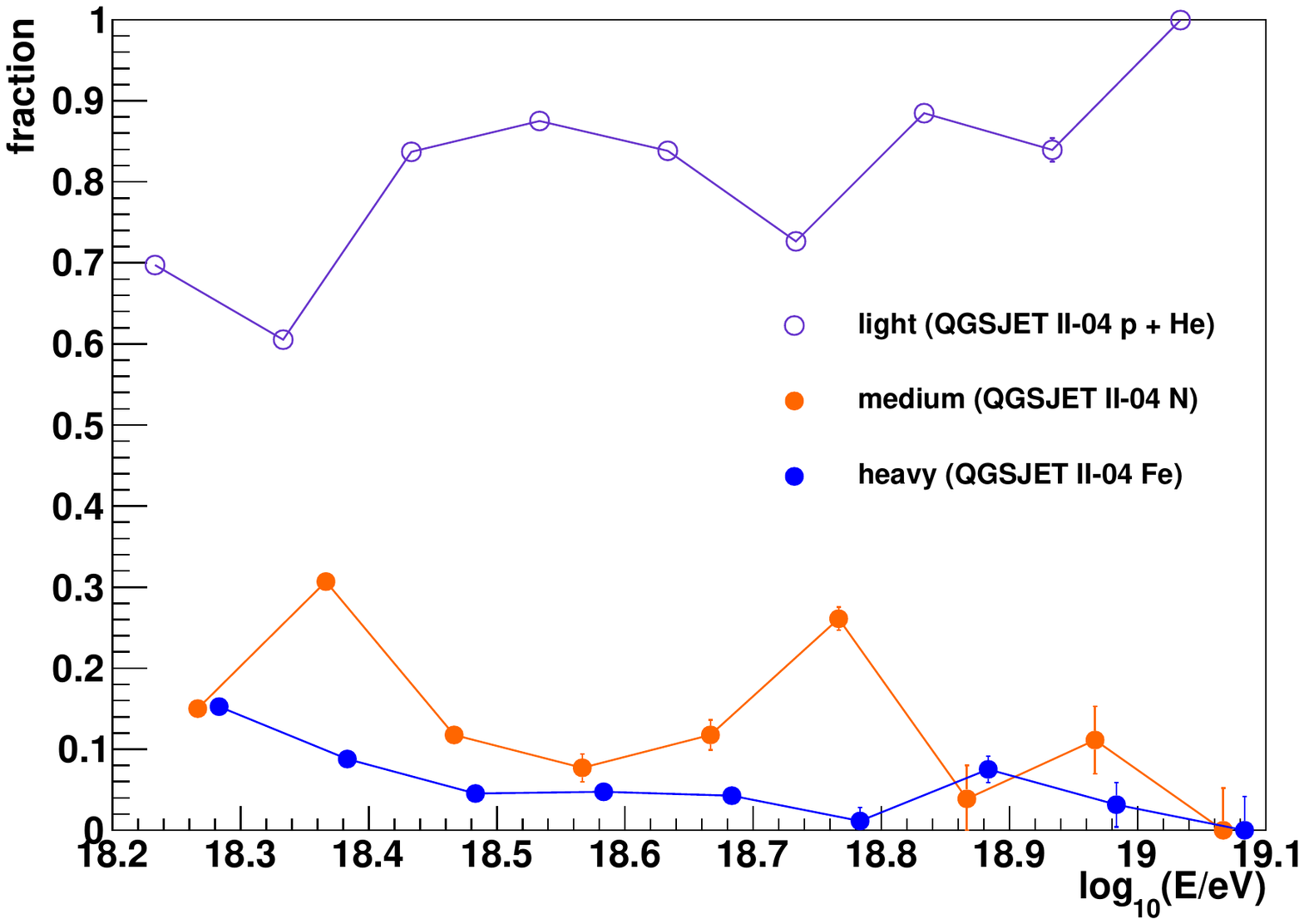}
  \caption{QGSJET~II-04 four component model fractions.}
  \label{fig:4comp_model_fractions}
\end{figure}

We find that in this mix model, correlations between the different
element fractions extracted from the fitter exist. In particular,
proton and helium are strongly correlated with $r < -0.9$ for nearly
the entire energy range. The least correlated elements are proton and
iron. This is due to the similarity of the proton and helium \xm{}
distributions. \mxm{} of helium and proton differ by only $~
25$~g/cm$^2$ and the \xm{} resolution of hybrid reconstruction is
about 18~g/cm$^2$. A future Monte Carlo study will investigate the
bias introduced into the fraction calculations that are caused by
these correlations. For this reason it is better to classify the
proton and helium contributions as ``light'' elements, with nitrogen
as ``medium'', and iron as
``heavy''. Figure~\ref{fig:4comp_model_fractions} shows the fractions
of these three categories of elements as a function of energy.

\section{Summary}\label{sec:summary}
Telescope Array has recently completed analysis of ten years of hybrid
\xm{} data. \mxm{} and \sxm{} of the data are consistent with
predominantly light elements such as QGSJET~II-04 proton and
helium. Tests of our \xm{} data against single elements require
systematic shifting of order the reconstruction resolution, but find
that the data is compatible with pure QGSJET~II-04 protons over the
entire energy range observed for that study ($10^{18.2}$ -
$10^{19.9}$~eV). If we extend comparisons of data to Monte Carlo
predictions by fitting multiple sources, we find agreement with
QGSJET~II-04 proton-iron mix, again with systematic shifting of the
data, resulting in a 95\% proton content. If we attempt to
simultaneously fit intermediate mass elements as well in a four
component mixture of proton, helium, nitrogen, and iron as Auger has
reported on, we can find good agreement with the data and the mix
without the need of systematic shifting of the data. This results in a
mixture of $\sim75$\% proton + helium, 17\% nitrogen, and 8\%
iron. Because we observe strong correlations in the fitting of
multiple components, particularly proton and helium, we report them as
a ``light'' component of the mixture.

\bibliographystyle{JHEP}
\bibliography{icrc2019_hanlon_596}

\end{document}